\documentclass[11pt]{article}

\usepackage[final]{acl}

\usepackage{times}
\usepackage{needspace}
\usepackage{float}
\usepackage{placeins}
\usepackage{latexsym}
\usepackage[T1]{fontenc}
\usepackage[utf8]{inputenc}
\usepackage{microtype}
\usepackage{booktabs}
\usepackage{amsmath,amssymb}
\usepackage{graphicx}
\IfFileExists{inconsolata.sty}{\usepackage{inconsolata}}{}

\title{Do Multilingual Encoders Produce Language-Consistent Semantic IDs?}

\hypersetup{
  pdftitle={Do Multilingual Encoders Produce Language-Consistent Semantic IDs?},
  pdfauthor={Abhinav Bohra, Anuj Bohra}
}

\author{
  Abhinav Bohra\thanks{Equal contribution.} \\
  Amazon \\
  \texttt{bohraabhinav8@gmail.com}
  \And
  Anuj Bohra\footnotemark[1] \\
  Rutgers University \\
  \texttt{anujbohra232003@gmail.com}
}

\begin{document}
\maketitle

\begin{abstract}
Semantic IDs (SIDs) compress item embeddings into discrete code sequences used in generative retrieval. We ask whether a multilingual encoder is sufficient for different-language renderings of the same product to receive language-consistent SIDs. Using Amazon ESCI listings rendered in English, Spanish, and Japanese, we test whether translations remain close to their English source, whether residual quantization is unusually sensitive to translation-induced movement, and whether multilingual or language-balanced quantizer fitting improves SID agreement. Multilingual E5 places translations measurably apart: under an English-heavy fit, a Japanese translation preserves the first SID code of its English counterpart in only 7.7\% of cases, compared with 89.0\% for an English rewording. Distance-matched product-directed controls produce nearly the same full-SID mismatch as translation, providing no evidence that the quantizer selectively amplifies language directions. Balancing the fitting mixture makes codebook use more uniform but further reduces cross-lingual prefix agreement: Spanish first-code consistency falls from 28.3\% to 6.6\%, while an English-only fit preserves it for 67.6\% of Spanish translations. These results show that multilingual exposure and balanced codebook use alone do not guarantee language-consistent SIDs.
\end{abstract}

\section{Introduction}

Multilingual item catalogs are not naturally aligned parallel datasets. The same product may be represented differently across languages because descriptions are written, translated, or localized through independent processes. A Semantic ID (SID) computed from listing text is therefore an identifier for a particular textual representation, not automatically for the underlying product. This distinction matters for multilingual retrieval and recommendation systems, where equivalent items must remain reachable across languages \citep{bonab2021xmrec}.

SIDs are often treated as language agnostic when they come from a shared multilingual encoder. This assumption depends on two conditions: translations of the same product must remain close in encoder space, and whatever displacement remains must not push the product into a different discrete code prefix. The distinction matters because similarity in a continuous embedding space decreases gradually, whereas exact SID lookup does not: one changed code can send the item to a different branch of the tree.

We test three questions: whether translations remain close to the English source relative to same-language rewordings, whether the quantizer is unusually sensitive to translation-induced movement, and whether multilingual or balanced quantizer fitting improves cross-lingual SID agreement. We use product listings from the Amazon Shopping Queries Dataset (ESCI) \citep{reddy2022esci}. Each listing is translated into Spanish and Japanese with NLLB \citep{nllb2022}, while English paraphrases provide a same-language reference. We compare translation-induced displacement with distance-matched isotropic and product-directed controls, and repeat the English--Spanish analysis with Qwen2.5-3B-Instruct to reduce dependence on a single translation model.

Multilingual E5 already separates equivalent products by language. Residual quantization \citep{lee2022rqvae, rajput2023tiger, vats2025comprehensive, vats2025systematic} turns this displacement into SID-prefix changes, but the product-directed control provides no evidence that the quantizer is especially sensitive to language directions. Under an English-heavy fit, E5 also assigns substantially less first-level capacity to Spanish and Japanese. A balanced mixture reduces this capacity gap but produces even lower prefix agreement across translations. LaBSE does not show the same capacity imbalance, indicating that this effect depends on the encoder--quantizer combination.

Together, these results show that encoder displacement, quantizer behavior, and balanced codebook use are distinct issues. Making codebook use more uniform across languages does not guarantee consistent SID prefixes, and multilingual exposure alone is insufficient for cross-lingual SID alignment.

\section{Related Work}
Residual quantization maps a continuous representation to a sequence of discrete codebook entries \citep{lee2022rqvae}. In recommendation, TIGER \citep{rajput2023tiger} used this idea to create semantic IDs for items and trained a sequence model to generate the ID of the next item.

Several recent papers approach multilingual alignment by changing how the identifier is built. MGR-CSC \citep{mgrcsc}, for example, represents
documents with multilingual keyword atoms and allows equivalent keywords to share atoms across languages. Parallel tokenizers \citep{paralleltok} work at the subword level instead, aligning independently trained vocabularies so that translated words can receive the same token indices. Both approaches build alignment into the identifier itself. We study the unmodified pipeline, where the SID is whatever the quantizer produces from a multilingual embedding, because that is the configuration practitioners deploy when they assume the encoder handles languages for free.

The closest work on quantizer stability is DRQ \citep{drqvae}, which studies how semantic IDs change when item embeddings are moved and uses code overlap
and codebook capacity to evaluate the quantizer. We follow a similar idea, but we find that the direction of the movement matters. A random change can move an
embedding by the same distance as a translation while still taking it through a region that is not typical of real product embeddings. We therefore add a
product-directed control, where the movement follows a direction defined by another observed product embedding. This is useful because language
representations are often concentrated in a narrow part of the embedding space
\citep{ethayarajh2019contextual,gao2019degeneration}. We also compare multilingual E5 with LaBSE \citep{feng2022labse}, which is trained to place
translations close together in a shared embedding space.

\section{Experimental Design}
\label{sec:design}

\subsection{Dataset}
We use ESCI because it provides a large real-world product-search catalog and includes US, Spanish, and Japanese locales, which motivate our three-language setting. However, ESCI does not provide a parallel catalog with a reliable cross-locale mapping for the same product. Comparing native locale listings would mix language with product identity, catalog composition, writing style, and field completeness. We therefore sample English listings and generate parallel Spanish and Japanese versions of the same product text.

We filter the US catalog by removing exact normalized title duplicates and obtain about 1M candidates. From these, we sample 20K products using seed 42. Each input contains the title, brand, color, and the first 120 words of the bullet points. The mean input length is 607 characters.

\subsection{{Translations and paraphrases}}
The main multilingual dataset uses NLLB-200-distilled-1.3B with greedy decoding for EN$\rightarrow$ES and EN$\rightarrow$JA. NLLB was run with a 128-token source and output cap. To measure the effect of source truncation, we reconstruct the exact truncated English input seen by NLLB and embed it as \textsc{EN-Trunc}. To reduce dependence on a single translation model, we also translate the full 20K corpus with Qwen2.5-3B-Instruct; this replication focuses on English--Spanish.

English paraphrases provide a same-language reference. The light paraphrase applies sentence reordering and a small set of synonym substitutions. Another structural transformation was also tested but produces a smaller embedding displacement and is included only in the appendix. These rule-based paraphrases are controls, not substitutes for naturally written paraphrases.

\subsection{{Distance-matched controls}}
We use distance-matched controls to test whether translation-induced movement is unusually disruptive to SID assignment. If translation is uniquely amplified by the quantizer, it should produce more SID changes than equally distant realistic perturbations.

All source embeddings are $L_2$-normalized. For source $\mathbf{x}$ and target cosine distance $d$, let $\theta=\arccos(1-d)$. The isotropic control samples a unit vector $\mathbf{u}$ orthogonal to $\mathbf{x}$:
\[
  \mathbf{x}_{\mathrm{iso}}
  =\cos\theta\,\mathbf{x}+\sin\theta\,\mathbf{u}.
\]
This control matches the translation distance but moves in an arbitrary direction in the embedding space.

For the product-directed control, we sample another product embedding
$\mathbf{z}$ and project its direction onto the tangent space at $\mathbf{x}$:
\[
  \mathbf{v}
  =\frac{\mathbf{z}-(\mathbf{x}^{\top}\mathbf{z})\mathbf{x}}
  {\left\|\mathbf{z}-(\mathbf{x}^{\top}\mathbf{z})\mathbf{x}\right\|_2},
  \qquad
  \mathbf{x}_{\mathrm{prod}}
  =\cos\theta\,\mathbf{x}+\sin\theta\,\mathbf{v}.
\]
This construction does not guarantee that every intermediate point lies on the true data manifold, but its direction is defined by an observed product embedding and is therefore more realistic than an arbitrary isotropic direction.

\subsection{{Encoders and quantizer}}
The main encoder is multilingual-E5-large \citep{wang2024me5}, a multilingual retrieval encoder, run in fp16 with 1024-dimensional outputs and the required \texttt{passage:} prefix. We use LaBSE \citep{feng2022labse}, which is explicitly trained for cross-lingual alignment, as an independent encoder check. Embeddings are $L_2$-normalized. The two embedding spaces differ sharply: E5 has a shared-offset proxy $\|\bar{\mathbf{x}}\|_2^2$ of 75.4\% and an average random-pair cosine similarity of $0.754\pm0.022$, compared with 32.6\% and 0.321 for LaBSE. 

The RQ-VAE uses three levels with 256 codes per level and a 256-dimensional latent space, Euclidean codebook assignment, and a cosine reconstruction loss. Inputs are centered before quantization and renormalized. Without centering and frequent dead-code recycling, the first level collapses to a single code (Appendix~\ref{app:collapse}). Remaining architecture and optimization settings are given in Appendix~\ref{app:compute}.

\subsection{{Fitting mixtures}}
We use three fitting mixtures to test whether multilingual exposure and language balance improve SID consistency. Each quantizer is fitted on 16K products and evaluated on a 4K-product held-out set. Condition A is an English-only baseline. Condition B uses 80\% English, 10\% Spanish, and 10\% Japanese, representing an English-heavy multilingual fit. Condition C uses equal proportions of all three languages to test whether balancing the fitting data improves cross-lingual agreement.

Product identities and counts remain fixed; only the rendering assigned during fitting changes. We train three seeds, 42, 123, and 789. The Qwen robustness arm uses separate English--Spanish mixtures of 100/0, 90/10, and 50/50; these are kept separate because they are not numerically identical to the three-language NLLB conditions.

\subsection{{Evaluation Metrics}}
We report paired cosine distance, five-fold cross-validated language-probe accuracy, Jaccard@10, full SID mismatch, and effective capacity:
\[
\begin{aligned}
K_{\mathrm{eff}}(L) &= 2^{H_L},\\
H_L &= -\sum_c p_L(c)\log_2 p_L(c).
\end{aligned}
\]
where $p_L(c)$ is the fraction of items assigned to code $c$ at level $L$. Prefix consistency at depth $d$ is the fraction of alternate renderings whose first $d$ codes match those assigned to the corresponding English listing. This metric measures reachability under exact SID-prefix lookup; it is not an end-to-end generative-retrieval score.

\section{Results}

\Needspace{20\baselineskip}
\subsection{Encoder separation by language}
\label{sec:geometry}

\begin{table}[H]
\centering
\small
\begin{tabular}{lrr}
\toprule
& E5-large & LaBSE \\
\midrule
Shared-offset proxy & 75.4\% & 32.6\% \\
Random-pair cosine & $0.754\pm0.022$ & 0.321 \\
\midrule
$d$(EN, ES) & 0.081 & 0.145 \\
$d$(EN, JA) & 0.113 & 0.165 \\
ES / paraphrase & 7.4$\times$ & 13.7$\times$ \\
JA / paraphrase & 10.3$\times$ & 15.6$\times$ \\
\midrule
Language probe & $99.9\pm0.1\%$ & 84.6\% \\
Jaccard@10 (EN, ES) & $0.41\pm0.21$ & $0.43\pm0.21$ \\
Jaccard@10 (EN, JA) & $0.33\pm0.17$ & $0.39\pm0.19$ \\
\bottomrule
\end{tabular}
\caption{Encoder diagnostics on 20{,}000 parallel products. The table uses the
original full-English source for comparability across encoders. Probe chance is
33.3\%.}
\label{tab:geometry}
\end{table}

For E5, the Spanish and Japanese translations are 0.081 and 0.113 cosine
distance from the English source, compared with 0.011 for the light English
paraphrase. A linear classifier trained on the frozen embeddings reaches
$99.9\pm0.1\%$ accuracy when predicting the language. The nearest-neighbor
sets also change: Jaccard@10 is 0.41 for Spanish and 0.33 for Japanese. The
translations therefore do not remain in the same local neighborhood as the
English listing.



Language classification accuracy is lower for LaBSE, at 84.6\%, but its
translated pairs are not closer in cosine distance. The two measures capture
different properties. Classification accuracy indicates whether language can
be recovered from the embedding population, whereas paired distance measures
whether two translations of the same product remain close.


\subsection{{Encoder displacement as prefix failure}}
\label{sec:prefix}

\begin{table}[!ht]
\centering
\footnotesize
\begin{tabular}{llrrr}
\toprule
Mix & Query translation & $d{=}1$ & $d{=}2$ & $d{=}3$ \\
\midrule
A & EN rewording & .901 & .619 & .396 \\
A & ES & .676 & .239 & .081 \\
A & JA & .643 & .171 & .037 \\
\midrule
B & EN rewording & .890 & .605 & .379 \\
B & ES & .283 & .055 & .016 \\
B & JA & .077 & .002 & .000 \\
\midrule
C & ES & .066 & .017 & .003 \\
C & JA & .048 & .009 & .001 \\
\bottomrule
\end{tabular}
\caption{E5 SID-prefix consistency on 4k held-out products, averaged over three seeds. A query succeeds at depth $d$ when its first $d$ codes match the SID assigned to the English listing.}
\label{tab:prefix}
\end{table}

Table~\ref{tab:prefix} measures whether translations preserve the SID prefix assigned to the corresponding English listing. Under the 80/10/10 mixture, first code consistency is 89.0\% for an English rewording, 28.3\% for Spanish, and 7.7\% for Japanese. Full SID agreement for Japanese is effectively zero. 

This is where the continuous encoder shift becomes a discrete prefix failure. A translated embedding may remain close to the English embedding in cosine distance while crossing the boundary of the first code. An exact prefix lookup then follows a different branch, and the English-indexed product is no longer contained in that subtree. 

\subsection{{Product-directed control}}
\label{sec:control}

\begin{table}[t]
\centering
\footnotesize
\setlength{\tabcolsep}{3pt}
\begin{tabular}{lrrr}
\toprule
\shortstack{Target\\distance} &
Translation &
Isotropic &
\shortstack{Product-\\directed} \\
\midrule
Spanish  & .981 & .756 & .976 \\
Japanese & .995 & .887 & .998 \\
\bottomrule
\end{tabular}
\caption{Full SID mismatch under the E5 80/10/10 quantizer, using
\textsc{EN-Trunc} as the source. Both controls match each translation's cosine
distance. The product-directed control uses a direction defined by another
observed product embedding.}
\label{tab:controls}
\end{table}

The isotropic control suggests stronger disruption from translation: full-SID
mismatch is 0.981 versus 0.756 for Spanish and 0.995 versus 0.887 for Japanese.
The product-directed control removes this gap, with 0.981 versus 0.976 for
Spanish and 0.995 versus 0.998 for Japanese. We therefore find no evidence that
the quantizer selectively amplifies language directions. Instead,
product-supported movements cross code boundaries more often than isotropic
ones. Because mismatch is already near saturation, small residual differences
may remain hidden, but the stronger control does not support the large effect
suggested by the isotropic comparison.

\Needspace{16\baselineskip}
\subsection{{Balanced fitting hurts prefix agreement}}
\label{sec:tradeoff}

\begin{table}[H]
\centering
\small
\setlength{\tabcolsep}{3pt}
\begin{tabular}{llrrrrr}
\toprule
Encoder & Mix & \multicolumn{3}{c}{$K_{\mathrm{eff}}$(L1)} &
\multicolumn{2}{c}{Prefix@1} \\
\cmidrule(lr){3-5}\cmidrule(lr){6-7}
& & EN & ES & JA & ES & JA \\
\midrule
E5 & B & 76 & 30 & 16 & .283 & .077 \\
E5 & C & 49 & 42 & 42 & .066 & .048 \\
LaBSE & B & 101 & 91 & 90 & -- & -- \\
\bottomrule
\end{tabular}
\caption{Effective first-level capacity on equal-size language pools and E5
first-code consistency with the English-indexed SID. Mixture B is 80/10/10;
mixture C uses equal thirds.}
\label{tab:tradeoff}
\end{table}

Under the E5 80/10/10 mixture, effective first-level capacity is 76/30/16 for
English, Spanish, and Japanese; equal-thirds fitting narrows this to 49/42/42.
Prefix agreement, however, falls from 28.3\% to 6.6\% for Spanish and from
7.7\% to 4.8\% for Japanese. The English-only quantizer preserves the first
code for 67.6\% of Spanish and 64.3\% of Japanese translations. Thus, balanced
code use does not imply cross-lingual code agreement. Qwen shows the same
trade-off: English--Spanish first-code agreement drops from 0.39 to 0.24 to
0.04 as the Spanish share of the fitting set increases
(Appendix~\ref{app:qwen}).

\section{Conclusion}
Cross-lingual SID inconsistency begins in the encoder space, but exact-prefix lookup makes the effect much more severe. A modest change in cosine distance
can alter the first code, after which the indexed product is no longer reachable through the same subtree. The product-directed control gives no
evidence that the quantizer is unusually sensitive to language directions. However, the quantizer still reflects the geometry of its inputs. Under an English-heavy fitting mixture, E5 assigns substantially less effective first-level capacity to Spanish and Japanese. Balancing the fitting mixture reduces this capacity gap, but it also lowers cross-lingual prefix agreement. By contrast, the English-only quantizer, despite seeing no non-English examples during fitting, preserves the first code for roughly two-thirds of the translations. These results suggest that multilingual exposure and balanced codebook use are not sufficient for SID alignment. Multilingual SID systems will likely need an objective that directly encourages parallel renderings to share assignments, rather than relying on a multilingual encoder or a balanced fitting corpus to produce that agreement indirectly.

\section*{Limitations}

The Spanish and Japanese texts are machine translations rather than naturally written listings because ESCI does not provide reliable cross-locale matches for the same products. Consequently, the measured displacement may reflect both language differences and artifacts introduced by machine translation. Replication with Qwen reduces dependence on a single translation model but does not eliminate this confound. The study also covers only Spanish and Japanese as non-English languages, limiting broader generalization. Finally, prefix consistency measures exact SID reachability, not end-to-end retrieval quality; the detailed Qwen analysis is limited to English--Spanish.

\section*{Ethics Statement}

The experiments use a public product-search dataset and contain no personal information. We measure differences between textual translations, not attributes of people. A failure to retrieve non-English listings could affect users in those locales, but this paper does not measure individual or population-level
harm.

\appendix

\Needspace{6\baselineskip}
\section{Level-1 Collapse and Mitigation}
\label{app:collapse}

Our initial RQ-VAE runs assigned every item to a single first-level code.
Later levels used roughly 45 codes each, reducing the observed ID space from
$256^3$ to approximately $45^2\approx2{,}000$ tuples. The training loss did
not clearly reveal this failure, but the resulting first-level drift and
utilization measurements were not meaningful.

\begin{table}[!htbp]
\centering
\small
\setlength{\tabcolsep}{3.5pt}
\begin{tabular}{llrrr}
\toprule
Variant & Change & L1 alive & $K_{\mathrm{eff}}$ & Collision \\
\midrule
V0 & baseline & 1 & 1 & 21\% \\
V1 & add centering & 22 & 20 & 1.9\% \\
V2 & RVQ library, cosine & 256 & 217 & 50\% \\
\textbf{V5} & \textbf{final configuration} & \textbf{256} &
\textbf{102} & \textbf{4.3\%} \\
V7 & reset every 2 epochs & 8 & 6 & 99\% \\
\bottomrule
\end{tabular}
\caption{Representative collapse ablations on 20{,}000 items. V5 is used in
the main experiments.}
\label{tab:collapse}
\end{table}

The dominant shared direction in the E5 embeddings places the raw vectors in
a narrow cone. Centering spreads the inputs presented to the quantizer, while
per-step dead-code recycling prevents one early code from dominating the
first level. Using cosine distance for codebook assignment kept more codes
active, but it also produced highly correlated levels and a large full SID
collision rate at this scale. The final configuration therefore uses
Euclidean assignment with a cosine reconstruction loss. These choices address
collapse in the present setting; they are not intended as evidence that the
same configuration is optimal for larger catalogs.

\FloatBarrier

\Needspace{6\baselineskip}
\section{Low Displacement Surface Perturbations}
\label{app:surface}

The rule based Spanish and Japanese perturbations move the E5 embeddings by
0.0064 and 0.0239, compared with 0.081 and 0.113 for the NLLB translations.
They produce little or no capacity asymmetry and substantially understate the
cross-lingual separation observed with NLLB. Their full SID mismatch remains
higher than that of the isotropic control, although neither condition is close
to saturation.

\begin{table}[!htbp]
\centering
\footnotesize
\setlength{\tabcolsep}{3.5pt}
\begin{tabular}{llrrrrrr}
\toprule
& & \multicolumn{3}{c}{Spanish} & \multicolumn{3}{c}{Japanese} \\
\cmidrule(lr){3-5}\cmidrule(lr){6-8}
Mix & Rendering & Trans. & Iso. & $\Delta$ & Trans. & Iso. & $\Delta$ \\
\midrule
B & surface & 42.8 & 24.7 & +18.1 & 77.7 & 47.2 & +30.5 \\
C & surface & 46.9 & 24.7 & +22.2 & 78.6 & 47.6 & +31.0 \\
B & NLLB & 99.0 & 82.0 & +17.0 & 99.8 & 91.4 & +8.4 \\
C & NLLB & 99.5 & 86.1 & +13.4 & 99.9 & 94.0 & +5.9 \\
\bottomrule
\end{tabular}
\caption{full SID mismatch against the original isotropic control. These
numbers are retained as measurements, not as evidence of language-selective
quantizer amplification.}
\label{tab:isotropic}
\end{table}

These results show why the realism of a displacement control must be considered
before interpreting its margin. The isotropic comparison alone cannot support
a language specific causal claim, because the product-directed control in
Table~\ref{tab:controls} removes nearly all of the observed difference.

\FloatBarrier

\Needspace{6\baselineskip}
\section{Qwen Full-Corpus Replication}
\label{app:qwen}

Qwen2.5-3B-Instruct translated the same 20{,}000-product corpus used in the main
study. The reported downstream comparison uses English and Spanish with
separate two-language mixtures: $A_Q$ is 100/0, $B_Q$ is 90/10, and $C_Q$ is
50/50. The NLLB values in Table~\ref{tab:qwen} are recomputed under the same
binary mixtures and should not be compared directly with the three-language
NLLB conditions in Table~\ref{tab:prefix}.

\begin{table}[!htbp]
\centering
\small
\begin{tabular}{lrr}
\toprule
Metric & Qwen & NLLB \\
\midrule
EN--ES distance & .0807 & .0760 \\
Distance / paraphrase & 9.9$\times$ & 9.3$\times$ \\
Jaccard@10 & .453 & .487 \\
$K_{\mathrm{eff}}$(L1), $B_Q$: EN/ES & 50/11 & 49/17 \\
Prefix@1 ES, $A_Q/B_Q/C_Q$ & .39/.24/.04 & .34/.13/.003 \\
Prefix@1 EN rewording & .78--.83 & .76--.83 \\
\bottomrule
\end{tabular}
\caption{Full-corpus English--Spanish translator replication. The two-language
mixtures are defined above and are separate from the main NLLB experiment.}
\label{tab:qwen}
\end{table}

Both translators exceed their isotropic control. Against the product-directed
control, the remaining margins are only 0.001--0.006 at a mismatch rate near
0.99. We do not treat those differences as evidence of selective amplification.

\FloatBarrier

\Needspace{6\baselineskip}
\section{Direction and Category Checks}
\label{app:robustness}

Back-translation produces high mismatch in both directions. In the reverse
setting, however, the Spanish and Japanese source texts are themselves machine
translations, and the audit uses only the isotropic control. We therefore use
this experiment only to rule out a simple one-way penalty against non-English
inputs.

\begin{table}[!htbp]
\centering
\small
\begin{tabular}{lrrr}
\toprule
Direction & Translation & Isotropic & Margin \\
\midrule
EN$\rightarrow$ES & .994 & .794 & +.200 \\
ES$\rightarrow$EN & .989 & .717 & +.272 \\
EN$\rightarrow$JA & .996 & .895 & +.100 \\
JA$\rightarrow$EN & .993 & .902 & +.090 \\
\bottomrule
\end{tabular}
\caption{Direction reversal under the E5 80/10/10 quantizer. The
product-directed interpretation from the main paper also applies here.}
\label{tab:direction}
\end{table}

We also divide the held-out English embeddings into 20 $k$-means clusters.
Under mixture B, Spanish full SID mismatch ranges from 0.968 to 0.997 across
clusters, with a standard deviation of 0.008. Japanese mismatch ranges from
0.998 to 1.000. The effect is therefore not limited to a single coarse
product group in the embedding space.

\FloatBarrier

\section{Compute and Reproducibility}
\label{app:compute}

The RQ-VAE encoder is $\mathrm{Linear}(1024,1024)$, GELU, LayerNorm,
$\mathrm{Linear}(1024,256)$, and the decoder mirrors it. The objective adds a
commitment term weighted by $\beta=0.25$ to the cosine reconstruction loss.
Each model trains for 20 epochs with batch size 1024 using Adam at a learning
rate of $10^{-3}$.

All experiments ran on a single NVIDIA L4 GPU with 24 GB of memory. NLLB
translation of 20{,}000 products into Spanish and Japanese took approximately
20 minutes and was cached. With translations and embeddings cached, the main
quantizer-training and analysis suite completed in approximately 3.3 minutes.
The implementation uses PyTorch, NumPy, sentence-transformers,
vector-quantize-pytorch, CTranslate2, Transformers, scikit-learn, and pandas.


\begin{thebibliography}{99}

\bibitem[Al Kautsar and Koto, 2025]{paralleltok}
Muhammad Dehan Al Kautsar and Fajri Koto. 2025.
Parallel tokenizers: Rethinking vocabulary design for cross-lingual transfer.
\emph{arXiv preprint arXiv:2510.06128}.

\bibitem[Bonab et~al., 2021]{bonab2021xmrec}
Hamed Bonab, Mohammad Aliannejadi, Ali Vardasbi, Evangelos Kanoulas, and James
Allan. 2021.
Cross-market product recommendation.
In \emph{Proceedings of CIKM}.

\bibitem[Ethayarajh, 2019]{ethayarajh2019contextual}
Kawin Ethayarajh. 2019.
How contextual are contextualized word representations? Comparing the geometry
of BERT, ELMo, and GPT-2.
In \emph{Proceedings of EMNLP-IJCNLP}.

\bibitem[Feng et~al., 2022]{feng2022labse}
Fangxiaoyu Feng, Yinfei Yang, Daniel Cer, Naveen Arivazhagan, and Wei Wang.
2022.
Language-agnostic BERT sentence embedding.
In \emph{Proceedings of ACL}.

\bibitem[Gao et~al., 2019]{gao2019degeneration}
Jun Gao, Di He, Xu Tan, Tao Qin, Liwei Wang, and Tie-Yan Liu. 2019.
Representation degeneration problem in training natural language generation
models.
In \emph{International Conference on Learning Representations}.

\bibitem[Huang et~al., 2025]{mgrcsc}
Yuxin Huang, Simeng Wu, Ran Song, Yan Xiang, Yantuan Xian, Shengxiang Gao, and
Zhengtao Yu. 2025.
Multilingual generative retrieval via cross-lingual semantic compression.
In \emph{Findings of EMNLP}.

\bibitem[Lee et~al., 2022]{lee2022rqvae}
Doyup Lee, Chiheon Kim, Saehoon Kim, Minsu Cho, and Wook-Shin Han. 2022.
Autoregressive image generation using residual quantization.
In \emph{Proceedings of CVPR}.

\bibitem[NLLB Team et~al., 2022]{nllb2022}
NLLB Team et~al. 2022.
No language left behind: Scaling human-centered machine translation.
\emph{arXiv preprint arXiv:2207.04672}.

\bibitem[Rajput et~al., 2023]{rajput2023tiger}
Shashank Rajput, Nikhil Mehta, Anima Singh, Raghunandan Keshavan, Trung Vu,
Lukasz Heldt, Lichan Hong, Yi Tay, Vinh~Q. Tran, Jonah Samost, Maciej Kula,
Ed~H. Chi, and Maheswaran Sathiamoorthy. 2023.
Recommender systems with generative retrieval.
In \emph{Advances in Neural Information Processing Systems}.

\bibitem[Reddy et~al., 2022]{reddy2022esci}
Chandan~K. Reddy, Llu\'is M\`arquez, Fran Valero, Nikhil Rao, Hugo Zaragoza,
Sambaran Bandyopadhyay, Arnab Biswas, Anlu Xing, and Karthik Subbian. 2022.
Shopping Queries Dataset: A large-scale ESCI benchmark for improving product
search.
\emph{arXiv preprint arXiv:2206.06588}.

\bibitem[Wang et~al., 2024]{wang2024me5}
Liang Wang, Nan Yang, Xiaolong Huang, Linjun Yang, Rangan Majumder, and Furu
Wei. 2024.
Multilingual E5 text embeddings: A technical report.
\emph{arXiv preprint arXiv:2402.05672}.

\bibitem[Wang et~al., 2026]{drqvae}
Xuesi Wang, Junjie Wang, Ziliang Wang, Weijie Bian, and Guanxing Zhang. 2026.
Decoupled residual quantization for robust semantic IDs in recommendation.
\emph{arXiv preprint arXiv:2606.01844}.

\bibitem[Yang et~al., 2024]{qwen25}
An Yang, Baosong Yang, Beichen Zhang, Binyuan Hui, Bo Zheng, Bowen Yu,
Chengyuan Li, Dayiheng Liu, Fei Huang, Haoran Wei, and others. 2024.
Qwen2.5 technical report.
\emph{arXiv preprint arXiv:2412.15115}.

\bibitem[Wang, 2023]{vqpytorch}
Phil Wang. 2023.
\emph{vector-quantize-pytorch}.
Software repository.


\bibitem[Vats et~al.(2025a)Vats, Raja, Kattamuri, and Bohra]{vats2025comprehensive}
Arpita Vats, Rahul Raja, Ashish Kattamuri, and Abhinav Bohra.
2025a.
\newblock Quantum natural language processing: A comprehensive survey of
models, architectures, and evaluation methods.
\newblock \emph{TechRxiv preprint}.
\newblock doi:10.36227/techrxiv.176238247.72060075/v1.


\bibitem[Vats et~al.(2025b)Vats, Raja, Kattamuri, and Bohra]{vats2025systematic}
Arpita Vats, Rahul Raja, Ashish Kattamuri, and Abhinav Bohra.
2025b.
\newblock A systematic survey of quantum natural language processing: Models,
encoding paradigms, and evaluation methods.
\newblock In \emph{Proceedings of the QuantumNLP: Integrating Quantum
Computing with Natural Language Processing}, pages 53--64.
Association for Computational Linguistics.
\newblock doi:10.18653/v1/2025.quantumnlp-1.8.




\end{thebibliography}
\end{document}